# Spitzer Publication Statistics

E. SCIRE[1], Luisa Rebull[1], Seppo Laine[1]

[1] IPAC, Mail Code 314-6, Caltech, 1200 E. California Blvd. Pasadena, CA 91125

Corresponding author: Elena Scire (escire@ipac.caltech.edu)

## Abstract

We present statistics on the number of refereed astronomy journal articles that used data from NASA's Spitzer Space Telescope through the end of the calendar year 2020. We discuss the various types of science programs and science categories that were used to collect data during the mission and discuss how operational changes brought on by the depletion of cryogen in May 2009, including the resulting budget cuts, impacted the publication rate. The post-cryogenic (warm) mission produced fewer papers than the cryogenic mission, but the percentage of the exposure time published did not appreciably change between the warm and cryogenic missions. This was mostly because in the warm mission the length of observations increased, so that each warm paper on average uses more data than the cryogenic papers. We also discuss the speed of publication, archival usage, and the tremendous efficacy of the Legacy and Exploration Science programs (large, coherent investigations), including the value of having well-advertised enhanced data products hosted in centralized archives. We also identify the observations that have been published the largest number of times, and sort them by a variety of metrics (including program type, instrument used, and observation length). Data that have the highest reuse rates in publications were taken early in the Spitzer mission, or belong to one of the large surveys (large either in number of objects, in number of hours observed, or in area covered on the sky). We also assess how often authors have cited the *Spitzer* fundamental papers or have correctly referenced the Spitzer data they used, finding that as many as 40% of papers have failed to cite the papers, and 15% have made it impossible to identify the data they used.

## 1 Introduction

Most major observatories track the number of refereed journal articles that use data from the given observatory. These paper number counts are frequently used to measure the scientific impact of an observatory (throughout this article, the word "paper" is used to mean a refereed journal article). The major observatories have broken down their publication counts by various metrics in an attempt to better understand how the observatory data were published (e.g., the *Chandra X-ray Observatory*, Blecksmith et al. 2005; the *Hubble Space Telescope*, Meylan et al. 2004). Cross-observatory comparisons (e.g., Crabtree 2008, Abt 2012) have also been attempted. These previous attempts have discussed the difficulty of defining a meaningful metric that quantifies the scientific impact of an observatory. Rots et al. (2012) proposed publication metrics based on the speed of publication, the fraction of the total observing time published, and the overall archival usage. These metrics are less sensitive to observatory-specific factors, such as the number of years the observatory has operated, the number of potential data users, the amount of research funding available, the breadth of the research projects, the number of observations that the observatory can make within a year, and the uniqueness of the data.

Publications that used data from the *Spitzer Space Telescope* (hereafter *Spitzer*) (Werner et al. 2004) can be used to study the merits of these various metrics due to the telescope's unique mission profile. We have divided the *Spitzer* science observations into two phases: the cryogenic mission and the warm mission, both of which used the same telescope and, for the Infrared Array Camera (IRAC; Fazio et al. 2004), the same instrument, but which have very different publication rates and counts. We find that we have to be careful when comparing the publications from *the same observatory* if the mission profile changes significantly over the lifetime of the mission.

Section 2 gives an overview of the *Spitzer* cryogenic and warm missions, including summary statistics of the number and the kind of observations taken, and of the data analysis funding. Section 3 discusses the methods used to find the publications. It compares full-text searches to just tracking the citations of the *Spitzer* fundamental papers and the official observatory acknowledgment statement. It also discusses the methods and accuracy of matching the papers to the data used in them. Section 4 addresses publication statistics, from the number of papers published to the percentage of the data published, and the speed of publication. It also includes a section on data reuse that looks at archival usage and observing programs whose data have the highest data reuse rates. Finally, there is a discussion of the number of refereed journal articles produced per hour of data observed (hereafter "papers per hour of observation") that compares the three *Spitzer* instruments, the observing program types and finally the observing program size and observation length.

The intent of this paper is to provide a coherent look at all the Spitzer publication statistics once the Spitzer project stopped counting publications in detail at the end of the mission (at the end of year 2020). Counting publications has been an ongoing project throughout the lifetime of the *Spitzer* mission and has been chronicled in conference proceedings. Scire et al. 2010 describes the design and implementation of the publications database, and presents some basic publication statistics. Since that time, the full text searches have been transitioned to use the Astrophysics Data System[1] (ADS) full text search engine. Scire 2014 had an initial look at why the publication rates changed so drastically between the cryogenic and warm missions after four years of warm operations. They concluded that the data were being published and that the larger program sizes had not slowed the publication of the data, but the average number of hours per paper had increased 3.5x between the warm and cryogenic missions. Scire 2018 provided a more in-depth look at the time between when the data are available in the archive and when they are published. It examined the difference between plotting this as hours of data and by individual observations. Some of the charts included in that paper have been updated here in Section 4.2 with an additional three years of data. Scire et al. 2020 provides an in-depth look at which programs and science categories produced the most papers and had the highest papers per hour of data rate. All of these papers provide earlier versions of the count of the total number of Spitzer papers, the percent of the total exposure time published, archival usage, and other publication statistics. The current paper expands upon Scire et al. 2020 by adding the final publications data from the year 2020 (22% of the papers from that year were published after those reported in Scire et al. 2020), expanding the analysis presented in the earlier conference proceedings, and including new analysis of the number of papers per hour of data rate for the different program types, program sizes and observation lengths.

## 2 The Spitzer Mission

The *Spitzer Space Telescope* was an infrared space-based observatory that was launched on 25 August 2003 and decommissioned on 30 January 2020 (6002 days). The portion of the mission in which science data were taken was split into two phases. The cryogenic mission ran from December 2003 (after the end of the in-orbit checkout and science verification period) until the exhaustion of the liquid helium coolant in May 2009 (5.5 years) during which all three of *Spitzer's* instruments were active (IRAC, Multiband Imaging Photometer for Spitzer or MIPS: Rieke et al. 2004, and Infrared Spectrograph or IRS: Houck et al. 2004). After the cryogen was exhausted, the *Spitzer* Warm and Beyond missions (hereafter collectively referred to as the "warm mission") ran from July 2009 until the decommissioning on 30 January 2020 (10.5 years). During the warm mission, the passive cooling on the telescope and instrument chamber was good enough to keep the spacecraft and telescope temperatures low enough for the operation of the two shortest wavelength (3.6 and 4.5 micron) channels of the IRAC instrument with essentially no loss in sensitivity (Carey et al. 2010). However, the temperatures were not cold enough to operate MIPS or IRS. With only one instrument warm mission operations were significantly different from the cryogenic operations. The change with the most impact was the reduction of the warm mission budget to less than one-third of that of the cryogenic mission budget, resulting also in the decrease of data analysis funding for observers. Due to these changes, larger observing programs that could span multiple years were allowed. With more time allocated to the larger programs, the number of programs approved in each cycle of proposals in the warm mission decreased (Storrie-Lombardi & Dodd 2010). Table 1 summarizes the main differences between the cryogenic and the warm missions.

---

[1] Astrophysics Data System; https://ui.adsabs.harvard.edu/

Table 1. Summary of the differences between the *Spitzer* cryogenic and warm missions. Portions of this table are reproduced from Scire et al. 2020 with permission from the author.

| | Cryogenic Mission | | | Warm Mission |
|---|---|---|---|---|
| **# of programs per observing cycle** | ~300 | | | ~60 |
| **Largest program** | 868 hours | | | 5286 hours (0.6 years) |
| **PIs** | 747 scientists from 38 countries | | | 335 scientists from 22 countries |
| **Total Data Analysis Funding ($K)** | 106560.1 | | | 37701.6 |
| | | | | (35% of cryogenic mission, spread over more hours of data) |
| **Most common proprietary periods of science observations (% of total time observed)** | 0 days: 29.4% | | | 0 days: 71.0% |
| | 90 days: 0.7% | | | 90 days: 5.9% |
| | 365 days: 68.0% | | | 365 days: 20% |
| **Hours of Science Observations** | 36534 hours | | | 81188 hours |
| **Instrument** | MIPS | IRS | IRAC | IRAC |
| **Wavelengths** | 24, 70, 160 micron imaging, SED mode | 5.2 to 38 micron spectroscopy, 16 and 22 micron imaging | 3.6, 4.5, 5.8, 8.0 micron imaging | 3.6, 4.5 micron imaging |
| **Hours of science observations** | 11717.9 | 14909.8 | 9906.0 | 81188.3 |

| % of hours of data taken | 32.1 % of the cryogenic mission | 40.8% of the cryogenic mission | 27.1% of the cryogenic mission | 100% of the warm mission |
|---|---|---|---|---|
| **Avg observation length** | 41.1 min | 46.2 min | 34.0 min | 44.7 min |
| **Number of hours in Legacy, Exploration Science, and Frontier Legacy (large or multi-cycle) programs** | 3724.1 | 1436.0 | 3612.5 | 45978.6 |
| **Number of GO, GTO, DDT, Snapshot (one cycle) programs** | 7993.9 | 13473.8 | 6293.5 | 35209.8 |
| **Papers published that used data from this instrument as of 08 Feb 2020** | 4414 | 1927 | 4529 | 1756 |
| **Average number of refereed journal articles per hour of data (see Section 4.4.1)** | 0.13 | 0.38 | 0.46 | for observations before cycle 12: 0.07 |

Over the lifetime of the *Spitzer* mission the science observations fell into a variety of categories (Figure 1). The telescope and instrument primary investigators were allocated 20 percent of the available observing time for the first 2.5 years of the science mission, and 15 percent thereafter until the end of the cryogenic mission, as part of the Guaranteed Time Observations (GTO) program. Director's Discretionary Time (DDT) proposals were submitted on a rolling basis throughout the mission and were approved by the Spitzer Science Center Director. All other time on the telescope was solicited through calls for proposals to the general astronomical community and competed through a peer-review process (this includes all General Observations (GO) time, Legacy, Exploration Science, Frontier Legacy and Snapshot programs[2]). The various program types and the changes the science program went though over the lifetime of the mission in response to the operating constraints are summarized below.[3]

---

[2] For exact solicitations of observing time please see the *Spitzer* Call for Proposals for all Cycles at https://irsa.ipac.caltech.edu/data/SPITZER/docs/spitzermission/observingprograms/proposalcycles/callsforproposals/

[3] For more information on all the programs selected for the Spitzer mission and the observing logs see https://irsa.ipac.caltech.edu/data/SPITZER/docs/spitzermission/observingprograms/

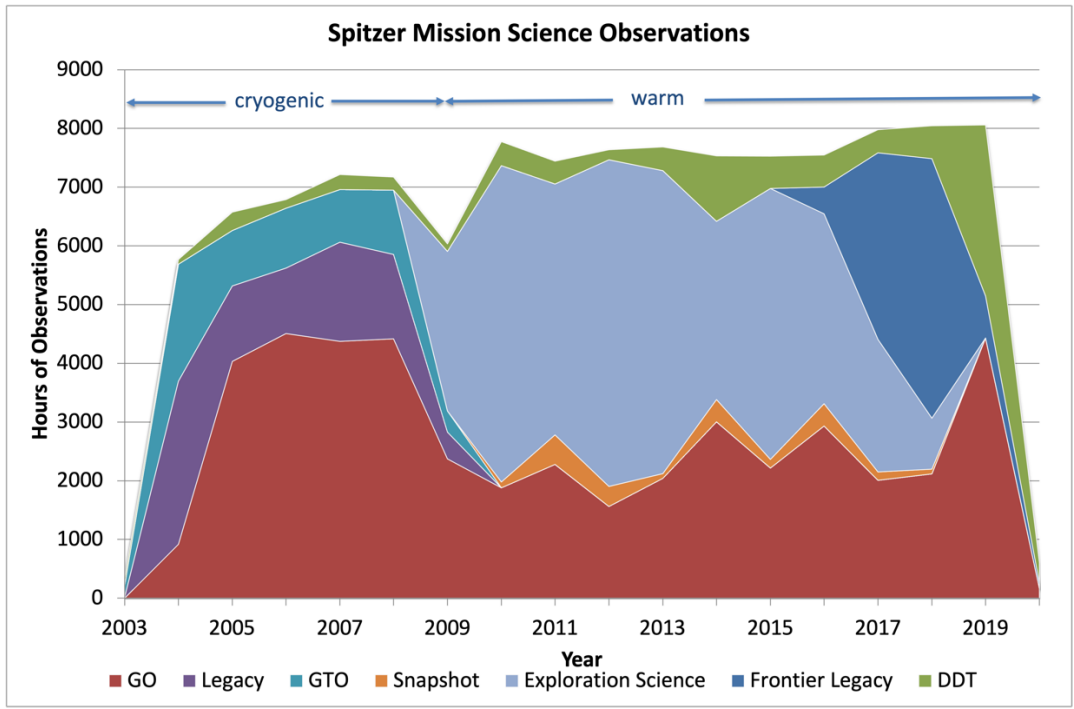

Figure 1: The program types for the science observations taken during the *Spitzer* mission. The cryogenic mission ran from Dec 2003 to May 2009, and the warm mission ran from July 2009 to Jan 2020. Legacy and GTO programs were in the cryogenic mission only. Exploration Science, Frontier Legacy, and Snapshot programs were in the warm mission only. It took 2–3 years to complete observations for the largest Exploration Science and Frontier Legacy programs. The dip in hours of science observations in 2009 was due to the IRAC Warm Instrument Characterization (IWIC) period between the cryogenic and warm missions. See the text for discussion.

Conceived before the launch, the initial Legacy[4] programs were large, coherent programs designed to execute early in the mission in order to provide a lasting legacy value for the *Spitzer* mission even if it ended early. The data taken in the Legacy programs had no proprietary period, and the science teams were contracted to produce enhanced data products (mosaics and/or catalogs) quickly. These were subsequently served to the community by the Spitzer Science Center and IRSA (NASA/IPAC Infrared Science Archive[5]). In the warm mission, the maximum program size was increased, and multi-year Exploration Science programs were introduced (similar to the Legacy programs, they had a minimum program size of 500 hours and would be executed over two years). Cycle 13 introduced Frontier Legacy programs, which were Exploration Science programs with even larger allocations (the minimum size for a Frontier Legacy program was 2000 hours of observing time)[6]. In total, 32 Legacy programs totaling 8691.4 hours were selected, with the largest program having 868 hours; 42 Exploration Science programs totaling 38412.4 hours and three Frontier Legacy programs totaling 8545.1 hours were selected. The largest Frontier Legacy program was allocated 5286 hours and took three years to complete, with the final observations occurring the week before decommissioning. There were almost as many hours in the three Frontier Legacy programs as in all the cryogenic Legacy programs combined. The data from the Frontier Legacy programs were taken late in the mission, and as of the start of 2020, no papers have yet been published from those data (which is not surprising: see section 4.2 below). The large number of hours of observations of Legacy/Exploration Science/Frontier Legacy programs (Figure 1) means that in the cryogenic mission 29.4% of the hours of science observations had no proprietary period. In the warm mission this number increased to 71% (Table 1).

---

[4] For a list of approved Spitzer Legacy programs and links to their abstracts and data products please see https://irsa.ipac.caltech.edu/data/SPITZER/docs/spitzermission/observingprograms/legacy/
[5] https://irsa.ipac.caltech.edu
[6] For a list of approved Exploration Science and Frontier Legacy programs and links to their abstracts and data products please see https://irsa.ipac.caltech.edu/data/SPITZER/docs/spitzermission/observingprograms/es/

Spitzer was designed to operate with only one instrument at a time to simplify operations and provide more robust power and thermal margins, so the observations were organized into instrument campaigns of $1-2$ weeks in length. Downlinks occurred once every ~12 hours. These two operational constraints, combined with the high proposal pressure and the small average allocation per observing program, broke up the GO time on *Spitzer* into many small programs. Cryogenic *Spitzer* also had GTO and DDT programs (Figure 1).

In the warm mission, only the two shortest wavelength channels on the IRAC instrument were operating. The reduced data volumes meant that the downlinking frequency (and the overall time spent downlinking) could be reduced. At the start of the warm mission downlinks typically occurred once every ~24 hours, and in the final years of the mission it was typical to have two to three days between downlinks. As less time was spent downlinking and swapping between instruments, there was an increase in the number of hours of observations per year (Figure 1). The change in the maximum program size led to a drop in the average number of observing programs per cycle from the cryogenic mission level of ~300 programs/cycle to ~60 programs/cycle in the warm mission (Table 1). In addition, it turned out that a large fraction of the science selected in the warm mission was time domain astronomy (e.g., exoplanet and brown dwarf light curves and exoplanet transit observations, or multi-epoch observations). The net effect of all these changes was that the warm mission had larger programs with more complicated constraints, some of which could span a few years. To help cope with the increasing complexity of scheduling the observations, the *Spitzer* project introduced a new program type. Snapshot programs were introduced in the second proposing cycle in the warm mission (Cycle 7), and had observations that were less than one hour in duration, with no constraints and low data volumes, and that allowed the science goals to be accomplished with only ~50% of the specified observations taken. They were designed to fill in holes around the heavily constrained exoplanet and other time series observations that dominated the warm mission.

The increase in DDT observations in 2014 (Figure 1) was due to the observations of the Frontier Fields (Soifer & Capak 2013a, 2013b, 2013c, 2013d, 2014a, 2014b). The Frontier Fields were a *Spitzer* and *Hubble Space Telescope* Director's Discretionary Time program to produce deep fields of strong lensing galaxy clusters[7]. At the end of the mission there were several small DDT proposal calls at regular intervals to fill the remaining observing time. The increase in GO time in 2019 took place in observing Cycle 14 that was limited to GO and Snapshot observations (there were no Exploration Science or Frontier Legacy programs solicited in that observing cycle).

In total, there were 2422 observing programs over the lifetime of the *Spitzer* mission, resulting in ~ 146000 science observations and 31630 calibration observations. During the cryogenic mission, 82% of all science time was available to the general community through competitive proposals; during the warm mission this fraction was 100%.

---

[7] https://irsa.ipac.caltech.edu/data/SPITZER/Frontier/overview.html

# 3 Publication Data Compilation
## 3.1 The Spitzer Bibliographical Database

Throughout the mission, the Spitzer Science Center maintained a bibliographical database[8] of refereed journal articles in which data from the observatory was used directly (rather than citing another paper or papers where such an analysis was done). It included papers that used the enhanced data products produced by the Legacy and Exploration Science teams and hosted at IRSA[9]. The first author of this article inspected every paper before adding it to the database.

All the numbers in this article were current as of 08 Feb 2021 (unless labeled otherwise) and include all the papers published through the end of the year 2020. At that time, the database contained 9304 papers, and there were 475,142 citations to those papers. The h-index was 248 (as calculated by ADS (using Hirsch's h-index). There were ~ 4400 unique first authors in those papers, and ~15,500 unique authors overall. In early 2021, the publications database grew at a rate of ~1.5 papers/day. For more information about the publications database, please see Scire et al. 2010, Scire 2014, Scire 2018 and Scire et al. 2020.

### 3.1.1 Finding the Papers

To find every paper that used data from *Spitzer*, we primarily relied on full text searches in ADS that use the name of the observatory, instruments, Legacy, and/or Exploration Science programs as search keywords. ADS full text searches include the text of the acknowledgments section of papers and figure captions. In some cases, we also track those papers that cite certain *Spitzer* fundamental papers. Of the two methods, the full text search produces the best results, as it does not rely on authors to correctly cite fundamental papers. While the Spitzer Science Center requested that all authors who use *Spitzer* data include a standard acknowledgment statement, not all authors have complied. Figure 2 shows the fraction of papers that correctly cited one of the fundamental papers for the telescope, instrument, Legacy, and/or Exploration Science programs or uses the Spitzer acknowledgment statement, as a function of time since the beginning of the mission. It also includes data for the number of papers that used data from an instrument and also cited that instrument's fundamental paper. Soon after the launch (in 2004) the number of papers that correctly cited the fundamental papers was almost 85%. Between 2004 and 2012 this number declined, and since 2012, an average of only 62% of papers that use data from *Spitzer* cited one of these fundamental papers or used the acknowledgment statement correctly. When an observatory is new, most of the data use is by the proposing observers, who are explicitly told in their funding contracts to cite the fundamental papers and to use the observatory acknowledgment statement. As the archival use starts to grow, it becomes the responsibility of the authors to seek out this information when using the data, and that explains why the correct citation rate falls. Of the three instrument fundamental papers, IRS data users most often cite the IRS fundamental paper – but only 53% of the time (Scire et al. 2020; see Section 4.4.1 for discussion). We urge all the authors using data from any facility to correctly cite fundamental papers and/or use acknowledgments in their papers. Lack of proper citations/acknowledgments is not a problem unique to Spitzer; see Chen et al. 2021.

---

[8] Originally at http://sohelp2.ipac.caltech.edu/bibsearch/, the database contents were moved to https://irsa.ipac.caltech.edu/bibdata/bibliography_table.html in fall 2021. They are also available by searching on the 'Spitzer' bibgroup in ADS and limiting it to refereed articles.

[9] Only papers that reference the original enhanced products produced by the Legacy/Exploration Science Team are included (and flagged in the bibliographical database as using the enhanced products rather than Spitzer Science Center pipeline data). If those enhanced products were then included in a next generation catalog (as often happens with extragalactic deep fields), papers that reference the next generation catalog are not included.

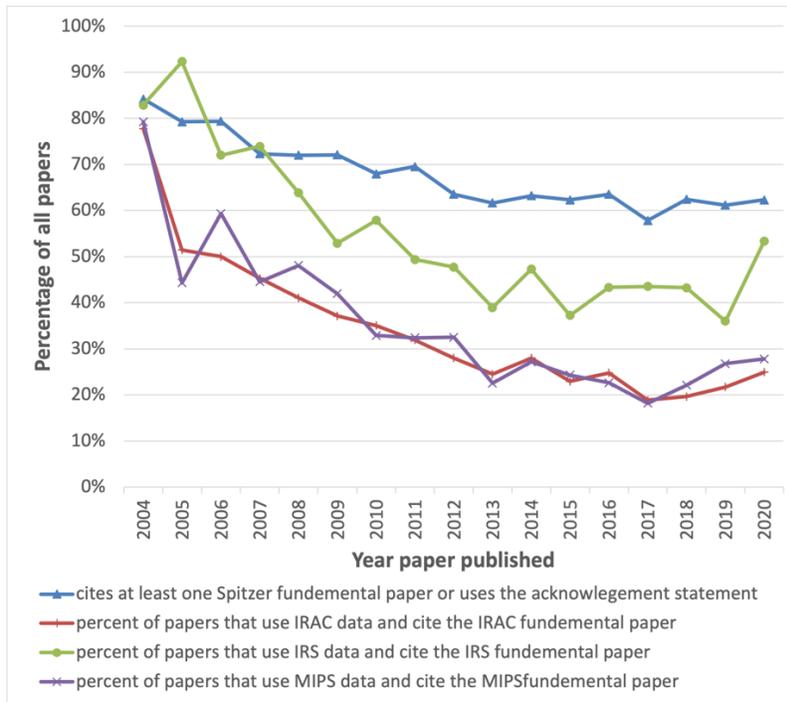

Figure 2: Blue line: Percentage of all papers that used data from *Spitzer* and that cited at least one of the telescope, instrument, Legacy, and/or Exploration Science fundamental papers or used the acknowledgment statement correctly. Since 2012, an average of 62% of papers that use data from *Spitzer* cited one of these items. Other lines, as shown: percentage of papers that use data from an instrument and also cite the instrument's fundamental paper.

Certain Legacy and Exploration Science programs have acronyms that are so common that they produce too many false hits in a full text search (e.g., SIMPLE: van Dokkum et al. 2005, SEDS: Fazio et al. 2008, etc.). For those programs, we checked every paper that cited the fundamental paper for those programs. We know that we are undercounting the number of papers that result from these programs, as the number of papers that cite the fundamental telescope and instrument papers is smaller than the number of papers based on the Spitzer telescope or the corresponding instruments (Figure 2 above and Table 2 of Scire et al. 2020). We conclude that future missions should come up with unique names for their fundamental programs so that an accurate publication tracking (even without citations) becomes feasible.

### 3.1.2 Data Matching

Each paper is matched to the data it uses by hand. Most papers include enough information to identify the data used to a relatively high degree of confidence, but approximately 15% of papers do not (Table 2). In those cases, the papers were linked to all the possible observations that could have been used or to no observations at all. This results in a slight inflation in the number of hours published in all the graphs that follow. We urge all authors to provide enough information in their papers to allow a unique identification of the data that were used (see also Chen et al. 2021).

Papers were categorized by how well the authors described the data used in them (Table 2). "Pretty Good" papers are ones where a search on the target returned a few (2–4) observations that could have been used, but the author did not provide enough information to narrow it down further. "Bad: linked to all data the authors had access to" category includes papers where the authors specified a target, but that target was observed many times throughout the mission, and it was not possible to narrow down which observation (or perhaps all of them) the author used, given the lack of information in the paper. In those cases, the paper was linked to all the possible observations to which the author had access to when the paper was written. The "Very Bad: Unlinked" category includes papers that definitely do use data from *Spitzer* but do not include even rudimentary information needed to identify the data used (a target list, which instrument was used, etc.). These papers are not matched to any data. "Other" papers reference an older paper when talking about which observations were used, so the match is only as good as the older paper. Table 2 only includes data from the years 2008 and 2020, because those are the only two years for which this classification is complete. As is the case with fundamental paper citations and acknowledgements above, authors' data description accuracy is slipping with time.

Table 2: Table showing how easy it was to match a published Spitzer paper to the Spitzer data that the paper is based on, in years 2008 and 2020, the only two years where this information is complete. Most papers include enough information to identify the data used to a relatively high degree of confidence ("Exact match", "Pretty good"). Approximately 15% of papers do not ("Bad", "Very Bad"). See the text for category descriptions. This table updates the data presented in Figure 2 in Scire et al. 2020. The final 22% of the papers from 2020 have been added to the totals here.

| Year | 2008 | 2020 |
|---|---|---|
| **Exact match** | 77.6% | 55.6% |
| **Pretty Good** | 9.9% | 20.4% |
| **Bad: Linked to all the data authors had access to** | 11.8% | 18.5% |
| **Very Bad: unlinked** | 0.2% | 1.5% |
| **Other** | 0.5% | 4.1% |

# 4  Results and Discussion
## 4.1  Number of Papers and Percentage of Exposure Time Published

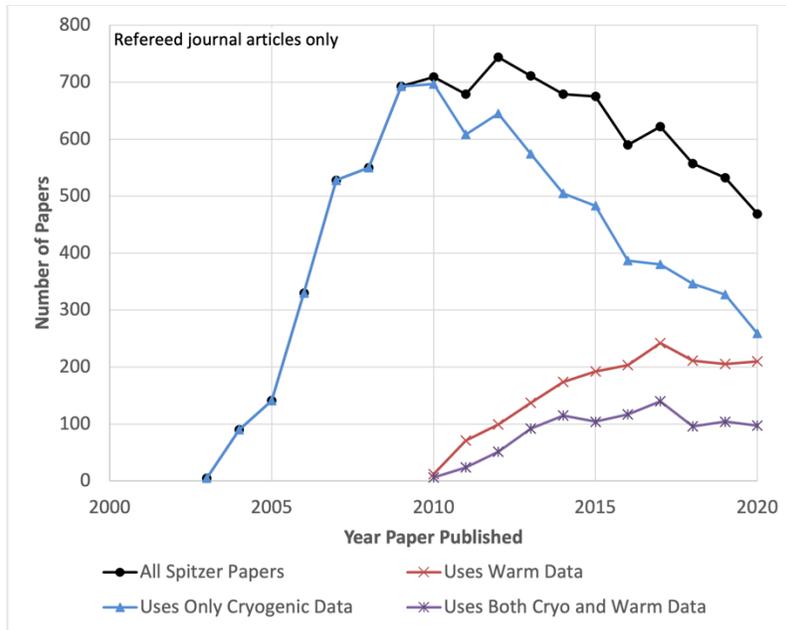

Figure 3: The number of refereed journal articles that used *Spitzer* data as a primary source per year (black line). The blue line represents the number of papers that used only cryogenic mission data taken before the depletion of the cryogen in May 2009, the red line represents the number of papers that used warm mission data taken after the depletion of the cryogen and the purple line represents papers that used both warm and cryogenic data. This plot includes papers that used the enhanced data products from the Legacy/Exploration Science programs.

Figure 3 shows a plot of the number of refereed journal articles that used *Spitzer* data as a function of time since the beginning of the *Spitzer* mission. The number of papers per year peaks in 2012 (744 papers), 2.5 years after the cryogenic mission ended, and then begins to decline (to only 469 papers in 2020). Figure 2 in Crabtree 2008 shows the rate at which various new ground-based telescopes begin to produce papers and concludes that a "plateau in productivity is reached seven to eight years after the initial publications." *Spitzer* reached a plateau in the number of papers five years after the initial publications, in 2009 and remained there though 2015. We believe the rapid increase in the number of *Spitzer* papers after launch was due to a combination of the well-funded GO and Legacy teams, and the large number of hours devoted to the Legacy programs, which had no proprietary period, early in the mission. The effects of archived, ready-to-use Legacy data products appear later than five years after launch (see section 4.4.3 below). The warm mission reached a rate of ~ 200 papers per year in 2015 and the rate has remained at that level. If the warm mission papers follow the same curve as the cryogenic ones, we would expect the publication rate to begin to drop in the year 2025 or 2026. We should note that 53% of the papers that use warm data also use cryogenic data. This is due to observers using the warm mission to expand sample sizes and area covered in surveys, as well as adding more epochs to time series. When they publish the new warm data, they also republish the old, cryogenic data.

Even though the telescope hardware remained the same and there was little loss in sensitivity in the two IRAC channels that remained operational in the warm mission, the number of papers produced from the observatory dropped from the cryogenic to the warm mission, despite the fact that the number of hours observed per year increased in the warm mission (Figure 1 and Table 1). Even though there are 2.2 times more hours of warm data than cryogenic data, there are significantly fewer warm mission papers. It should also be noted that after the cryogenic mission ended in 2009, the plateau of ~ 700 papers per year lasted until 2015, so it took six to seven years for the number of cryogenic papers to begin a slow decline. For both the warm and cryogenic missions, it took approximately five years after the start of the mission for the number of papers being published to reach a steady state.

Even though there are not as many warm papers as cryogenic ones, by examining the percentage of exposure time published (Figure 4, Table 3), it can be clearly seen that the warm data are getting published and used in several different investigations. Rots et al. (2012) argued that a simple count of the number of refereed journal articles per year is not a good metric for comparing the productivity between observatories, and proposed that the percentage of the exposure time published was a better metric. When comparing the *Spitzer* cryogenic and warm missions, this becomes very evident – the percentage of the exposure time published did not drop in the warm mission. Researchers were still using the data, just in very different ways than researchers did with the cryogenic data.

There are many related reasons that led to the decrease in the number of papers for *Spitzer*, and the sections that follow attempt to delve into the various reasons in an attempt to ascertain which ones are the most important in contributing to the decrease in the number of papers.

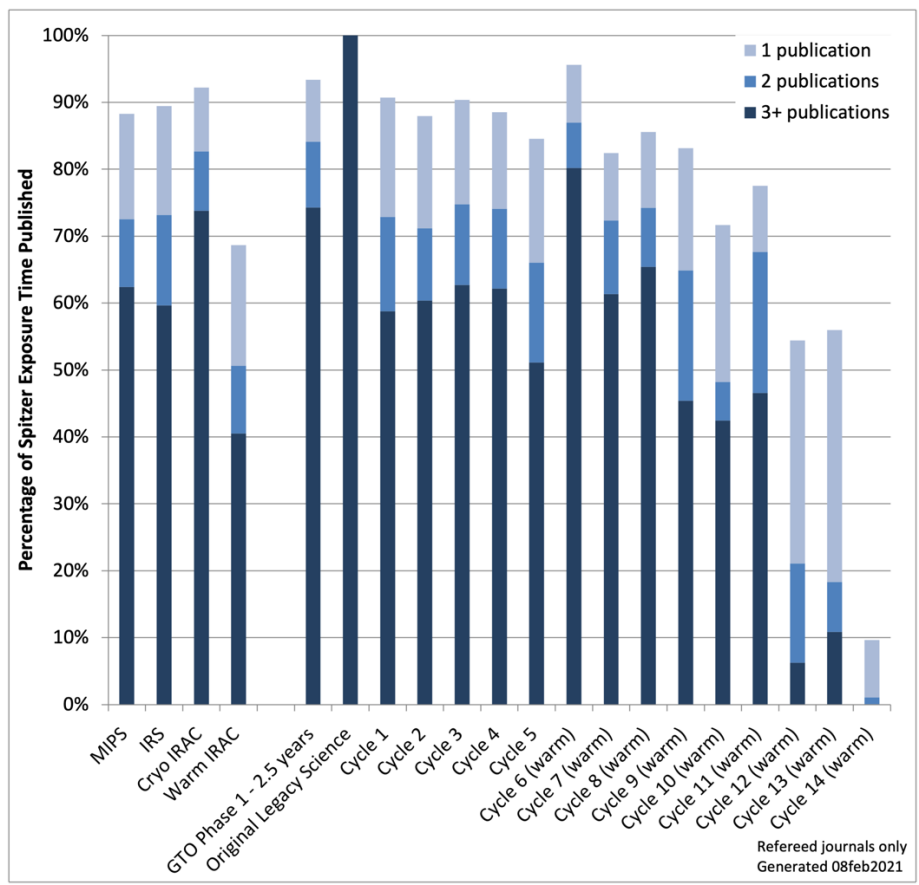

Figure 4: Percentage of the exposure time published, grouped by the instrument (left-hand side) or the observing cycle (right-hand side). Most data get published at least once (though there is a natural delay between the observation and the publication), and a significant fraction gets published many times. These numbers were calculated by checking how many times an observation had been published, then normalizing by the observation duration. Cycles refer to how the observations were broken up in the operations database. GTO phase 1, Original Legacy Science and Cycle 1 together made up the first ~ two years of the mission. All the program types, including GTO and Legacy, are included in the bars for Cycles 2 and up.  At the time of writing of this article, the data taken in Cycles 12, 13 and 14 were still relatively

new, and observers were still working on publishing these data. The last Cycle 11 observation occurred on 04 July 2017. A version of this figure was published as Figure 4 in Scire et al. 2020. Since then, the percentage of data published in Cycles 12, 13, and 14 has increased. The raw data for this plot are in Table 3

Table 3: Percentage of the exposure time published, grouped by the instrument or the observing cycle. The data in this table are plotted in Figure 4.

|  | Percent of Exposure Time Published at Least Once | Only 1 publication | Only 2 publications | 3+ publications |
|---|---|---|---|---|
| By Instrument | | | | |
| MIPS | 88.3% | 15.7% | 10.1% | 62.4% |
| IRS | 89.4% | 16.3% | 13.5% | 59.6% |
| Cryo IRAC | 92.2% | 9.5% | 8.9% | 73.8% |
| Warm IRAC | 68.7% | 18.0% | 10.1% | 40.5% |
| By Cycle | | | | |
| GTO Phase 1 - 2.5 years (cryo) | 93.4% | 9.2% | 9.9% | 74.3% |
| Original Legacy Science (cryo) | 100.0% | - | - | 100.0% |
| IOC/SV and Cals | 90.7% | 17.8% | 14.1% | 58.8% |
| Cycle 1 (cryo) | 88.0% | 16.8% | 10.8% | 60.4% |
| Cycle 2 (cryo) | 90.3% | 15.6% | 12.0% | 62.7% |
| Cycle 3 (cryo) | 88.5% | 14.4% | 11.9% | 62.2% |
| Cycle 4 (cryo) | 84.5% | 18.5% | 14.9% | 51.1% |
| Cycle 5 (cryo) | 95.6% | 8.6% | 6.8% | 80.2% |
| Cycle 7 (warm) | 82.4% | 10.1% | 11.0% | 61.3% |
| Cycle 8 (warm) | 85.5% | 11.3% | 8.8% | 65.4% |
| Cycle 9 (warm) | 83.2% | 18.3% | 19.5% | 45.4% |
| Cycle 10 (warm) | 71.6% | 23.4% | 5.8% | 42.5% |
| Cycle 11 (warm) | 77.5% | 9.9% | 21.0% | 46.6% |
| Cycle 12 (warm) | 54.4% | 33.3% | 14.9% | 6.2% |
| Cycle 13 (warm) | 56.0% | 37.7% | 7.5% | 10.8% |
| Cycle 14 (warm) | 9.7% | 8.6% | 1.0% | 0.1% |

For *Spitzer*, approximately 10% of the cryogenic science data taken remain unpublished over a decade after the data were taken (Figure 4, Table 3). A similar percentage of unpublished data (10%) can be seen for the *Chandra X-ray Observatory* (Rots et al. 2012; Figure 5). Rots et al. also find that for *Chandra* "as much as 60% of all exposure time may eventually be presented in publications more than twice." It might be advantageous for observatories to look into their unpublished data to try to determine why they remain unpublished, and to call attention to these unpublished data.

As the *Spitzer* publications became more archival and the data analysis funding granted to PIs decreased (Table 1), authors shifted to journals without author fees (such as the Monthly Notices of the Royal Astronomical Society; MNRAS). The share of all *Spitzer* papers published in MNRAS rose from 10.4% in 2009 (the final year of the cryogenic mission) to 28.0% in 2020 (the final year of the warm mission), while the share of the papers published in the Astrophysical Journal (ApJ) dropped from 56.1% to 32.2% over the same time period. Other major journals retained a relatively constant percentage of *Spitzer* papers over the same time frame.

## 4.2 Speed of Publication

A very important factor to consider when counting publications is the speed with which observers generate papers and transit the publications process. Any changes made to the mission profile will take time to propagate out into the publication statistics. For the Legacy and Exploration Science programs with hundreds or thousands of hours of data, there is also a lag introduced while the data are being acquired if the nature of the program is such that the intended science cannot be accomplished without all the data in hand. In Figure 5 we examine the time between when the data were observed and when they were published.

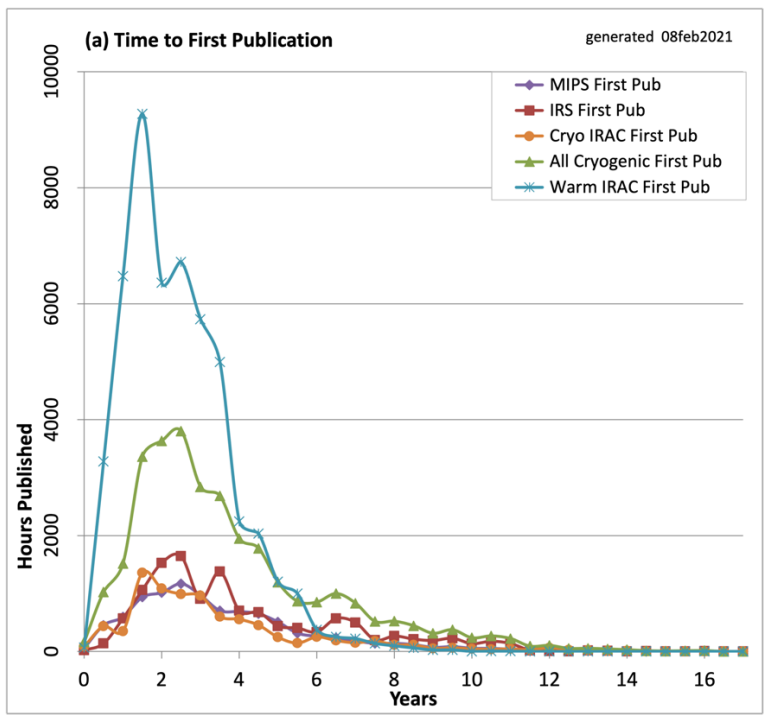

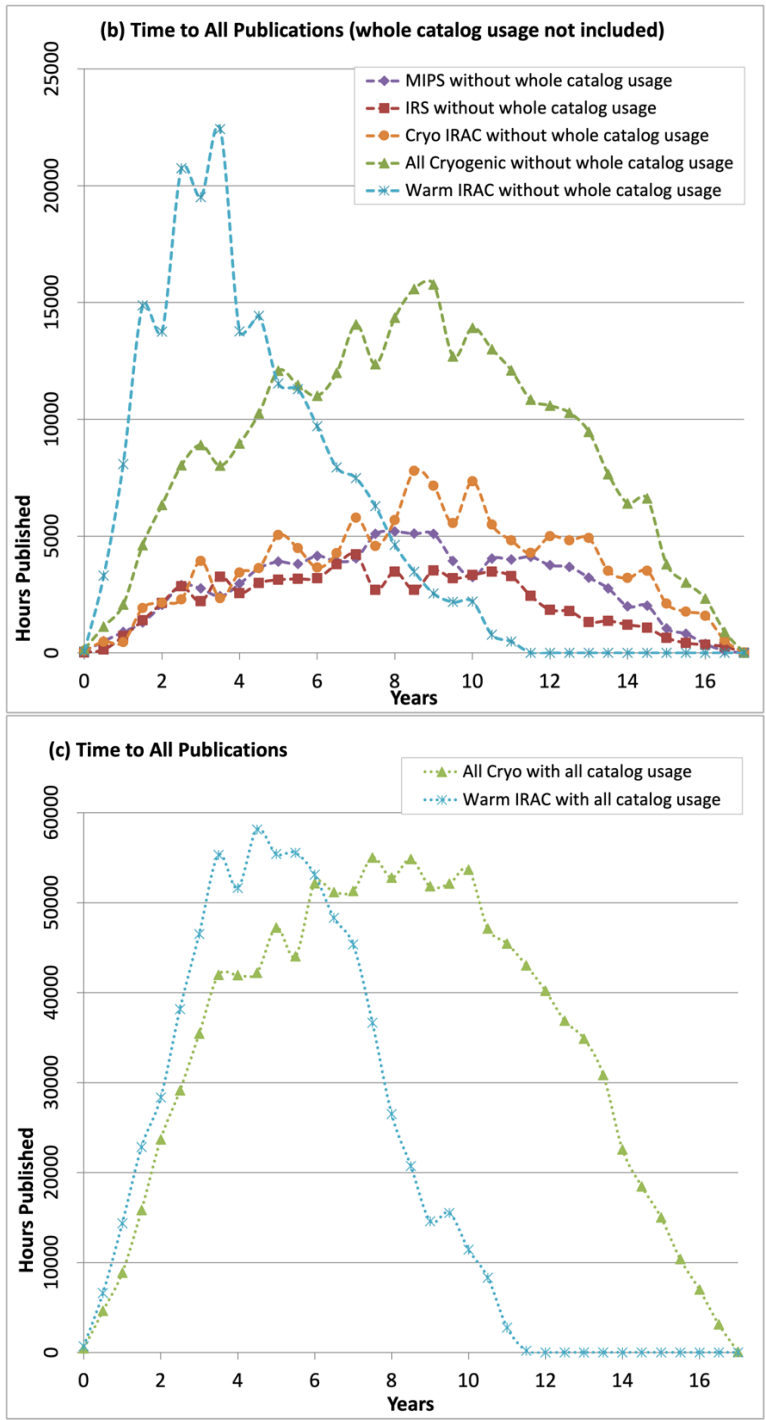

Figure 5. Time elapsed between the data becoming available in the archive to the PI (Principal Investigator) and (a) the first publication of the data and (b) and (c) all the publications of the data. (c) includes those instances where the paper used the entire catalog/mosaics from the Legacy/Exploration Science programs (see text), while (b) does not. The vertical axis is hours of published observations. The bin size is 6 months. Panels (a) and (b) include breakdowns by the mission phase (cryogenic, warm) and by the instrument. (c) only includes the mission phase. Panel (a) is in some sense a zoom of (b) which is a zoom of (c). MIPS, IRS and Cryo IRAC together make up the All Cryo line and ceased operations in May 2009 (11.75 years ago). Warm IRAC operated from July 2009 to Jan 2020 (11.5 years ago to 1 year ago).

See the text for discussion. A version of (a) and (b) appeared as Figures 3 and 4 of Scire (2018) They have been updated with an additional three years of publication data in this paper.

The mean time to the first publication of the data (Figure 5a) varies by instrument: data from the cryogenic IRAC instrument is published ~ 6 months faster than data from the MIPS and IRS instruments. Data from IRAC are considered easier to use and most closely resemble visual wavelength data with which many astronomers have experience with. Spectral data (IRS) require more data analysis and a deeper understanding of the physics involved, and far-infrared data from MIPS are generally nontrivial to reduce. During the cryogenic mission the three instruments operated in series, but during the warm mission only the IRAC instrument was operating. This meant that Warm IRAC was generating three times as many hours of data as any one of the cryogenic instruments in the same amount of time. In addition, the cryogenic instruments only operated for 5.5 years and Warm IRAC operated for 10.5 years. These two factors caused the Warm IRAC first publication curve to be much steeper than the Cryogenic IRAC, MIPS or IRS first publication curves, because it contained many more hours of data. Moreover, those warm IRAC data were very similar to the cryogenic IRAC data, so quite a bit of software was already written and ready to be used on the warm IRAC data as soon as it arrived on the ground.

There are several interesting things to note about the data as they are republished (Figure 5b). When the data are still relatively new (years 0 – 6) the three cryogenic instruments trend together very closely. As the data age, cryogenic IRAC data are more heavily reused than MIPS and IRS data. This is most likely due to the fact that in the warm mission, only the IRAC instrument was operating, and observers were expanding sample sizes and getting more epochs of data. When they publish the warm mission observations, they go back and reuse the complementary cryogenic data at the same time. The existence of the warm mission has increased the "value" of the cryogenic data. In addition, a surprising amount of data taken very early in the mission is still very heavily used (years 14 – 16). We note that IRAS data, taken in 1983, are still being frequently used in papers.[10] For more information on which data from early in the *Spitzer* mission are the most heavily republished, see Sections 2.2 (for specific programs) and 2.4 (for science categories) in Scire et al. 2020.

The "with whole catalog usage" and "without whole catalog usage" bins that are plotted separately in Figure 5 b) and c) denote how the data were used within a paper. A paper can only be in one of these categories. If the author states that they downloaded the SCOSMOS data (Sanders et al. 2005 and 2006) from the *Spitzer* archive and reduced them themselves, then they are in the "without whole catalog usage" bin. If an author made a statement that they used a fraction of the catalogue/mosaics, then they are also placed into the "without whole catalog usage" bin. This is very common when authors are using GLIMPSE data (Churchwell et al. 2004 and 2005). They will go into the mosaic of the entire Milky Way and retrieve a postage stamp of their object of interest. It is also common with the Spitzer Infrared Nearby Galaxy Survey (SINGS; Kennicutt et al. 2004) data; authors are far more likely to download a mosaic of one galaxy than to use the entire SINGS dataset.

If a paper included a statement along the lines of "We downloaded the SCOSMOS catalog/mosaics from IRSA" and then referenced the *Spitzer* fundamental paper for that program (Sanders et al. 2007), then the paper was marked as "with whole catalog usage." This is very common for extragalactic deep field surveys where most authors choose not to reduce the data themselves but rely upon catalogs/mosaics. If the authors instead cited a paper other than the *Spitzer* fundamental paper for that Legacy/Exploration Science program, then the paper was not included in the counts of *Spitzer* papers. Those cases are usually where someone incorporated the *Spitzer* catalog into a new catalog with additional data from another observatory. As such, the *Spitzer* data are no longer a primary source.

---

[10] Bibliography of recent refereed papers using popular datasets at IRSA:
https://irsa.ipac.caltech.edu/bibdata/bibliography_list.html

The number of hours of data published by papers that are using the entire catalogue/mosaics from the Legacy/Exploration Science programs (Figure 5c) far exceeds any other usage of the data. When plotted by the hours of observation, this usage is dominated by the investigators who use the catalogs or mosaics from the extragalactic field surveys. For example, each paper that uses the cryogenic SCOSMOS catalog is essentially using 627 hours of data. (The non-extragalactic surveys are also heavily used, but generally do not use the whole catalog at once; see Section 4.4.3.) The existence of the enhanced products, their clear documentation, and the ease with which people can find and use them, have all greatly increased the value of these data. Observatories looking to increase the number of publications or to maximize the scientific use of the data products should encourage observers (especially of large, coherent projects) to deliver ready-to-use data back to a publicly available archive, such as IRSA[11].

When compared with *Chandra* (Rots et al. 2012, Figures 1 and 3), the *Spitzer* curves in Figure 5 show similar shapes. They both show a ramp-up that takes ~2 years to the first publication when the data are new, and a long tail of older data being republished.

---

[11] https://irsa.ipac.caltech.edu/irsa-dataQA.html

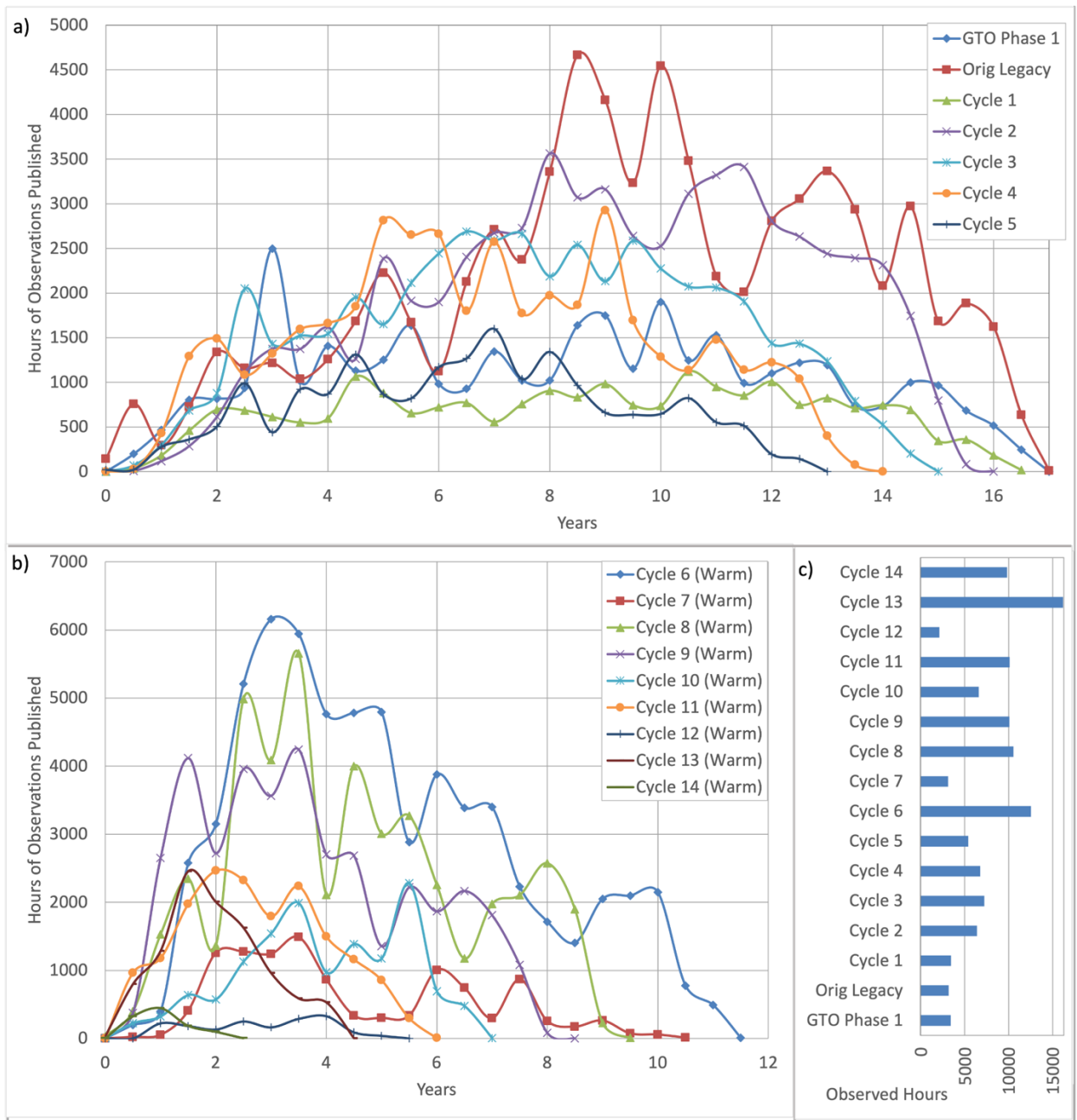

Figure 6: Time elapsed between when the data were released to the *Spitzer* archive, and all the publications of the data, sorted by the observing cycle for the (a) cryogenic and (b) warm missions. (c) Total number of hours of observations in each cycle. A few cycles were 1.5- or 2-year cycles (6, 11, 13) with a shorter cycle starting one year into them (7, 12). The cryogenic mission had many small programs producing papers at a constant rate; the warm era has more large programs that do not publish until all of their observations have been taken. An earlier version of this figure in a different format appeared as Figure 6 of Scire (2018). It has been updated with an additional three years of observation and publication data in this paper.

In a plot of the time needed to publish, sorted by the observing cycle in which the data were taken (Figure 6 a), the cryogenic mission Cycles 1 – 5 tend to trend together for the first 2-4 years after the data were taken. With ~300 programs per observing cycle in the cryogenic mission, any observer who is slow to publish generally only has a small number of hours of data, which does not materially impact the shapes of the curve in Figure 6. With only ~ 60 programs per observing cycle in the warm mission, a single slow observer is much more likely to have a more significant fraction of the total time observed, and their delay can really impact the shape of the curve. The time needed to publish data from the warm mission is much more variable – this is most likely due to the increase in the program size and the decrease in the overall number of programs per observing cycle. A few programs with thousands of hours of data took two or three years to observe. In those programs, observers are unlikely to use the data taken early in the observing cycle until they have received all of their data. This introduces an additional lag in the publication of data from these very large programs.

As the data age, some data from some observing cycles are more heavily reused than other data. GTO Phase 1, Cycle 1 and the Original Legacy cycles all contained approximately the same number of hours of observations and were all observed at the same time at the beginning of mission (Figure 6 a) and c)), but 16 years later the data from the Original Legacy programs are republished at a much higher rate than the GTO and Cycle 1 data. See Scire et al. 2020 for an in depth discussion of this topic.

## 4.3   Archival Data Reuse

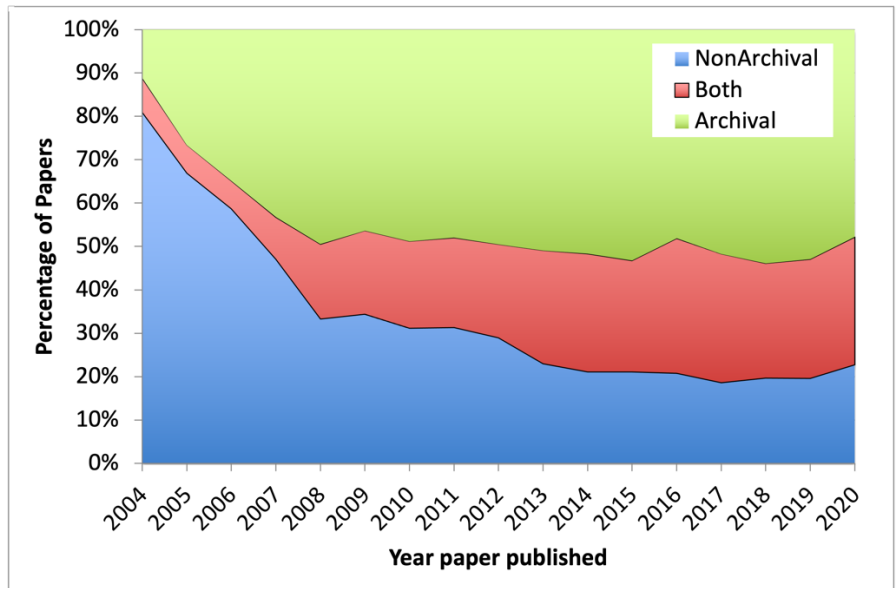

Figure 7: Archival *Spitzer* papers. To calculate this, the last names of the paper's authors were compared to the last names of all the people on an observing proposal. An archival paper is one where no authors match anybody on the proposal. A non-archival paper has at least one match for all the programs used in the paper. Papers classified as both use data from multiple programs and have at least one program that is archival, and one program that is non-archival. This graph does not take into account the amount of time that has passed between an observation was taken and when it was published. An earlier version of this figure appeared as Figure 10 in Scire (2018). It has been updated with an additional three years of publication data here.

Most of the papers that used data from *Spitzer* have been archival papers or used a mix of proprietary and archival data (Figure 7). An archival paper is determined by comparing the paper authors' last names with the last names of people listed on the observing proposal for all the data are used in the paper. The decision pre-launch to have the Legacy programs with no proprietary period (29.4% of the cryogenic data had no proprietary period; Table 1) and to make available ready-to-use enhanced products led to a rapid increase in the number of papers using Legacy data early in the mission and increased the fraction of archival papers produced. The warm mission had even less proprietary data (71% of warm data had no proprietary period). It is very important to have a well-advertised and easy-to-use archive with good documentation; as a mission ages, a large fraction of the data usage comes from people unaffiliated with the original observing program. Meylen et al. 2004 calculated that 34% of *Hubble* papers were archival, based on a comparison of the paper authors with PIs (primary investigators) or CoIs (co-investigators) on the programs. *Spitzer's* fraction of archival papers appears to be higher, perhaps due to the large amount of data with no proprietary period, although it would be interesting to see how *Hubble's* archival fraction has changed since 2004.

Information about which program have the heaviest data usage rates has been previously published in Section 2.2 in Scire et al. 2020 and discussed at length there. In brief, Scire et al. 2020 found that the programs with the heaviest data reuse rate when calculated by the number of hours published divided by the number of hours observed tend to be dominated by the Legacy or Exploration Science surveys of extragalactic survey fields (SCOSMOS: Sanders et al. 2005 and 2006, GOODS: Dickinson et al. 2004, etc.). This is due to the fact that when authors use the extragalactic survey field data, they use all of it at once to reach the required depth. Other fields on the sky that have large numbers of hours published relative to the number of hours observed are the galactic plane (Churchwell et al. 2004 and 2005, Benjamin et al. 2006, Carey et al. 2006), the Large and Small Magellanic clouds (Meixner et al. 2005 and Gordon et al. 2007) and three star forming regions (Evans et al. 2004). However, if one plotted the programs that have the largest number of papers (Scire et al. 2020 Figure 6), then the shallower surveys of many astrophysical objects dominate, with GLIMSPSE I (Churchwell et al. 2004) and the Spitzer Infrared Nearby Galaxy Survey (SINGS; Kennicutt et al. 2004) producing the most papers. In addition, some Guaranteed Time Observer (GTO) programs have extremely large numbers of per hour of observation rate because the GTO programs selected their targets first before the launch. That had the effect of concentrating many of the interesting and easy observations of small fields into GTO Phase 1

## 4.4 Number of Published Papers per Hour of Observation

Here we address the number of published papers per hour of observation broken down by the instrument, the observing program category (with special attention paid to the Legacy and Exploration Science Programs) and the program allocation and average observation length. For information about the papers per hour of observation rate for the various science categories, see Section 2.4 in Scire et al. 2020.

### 4.4.1 Instruments

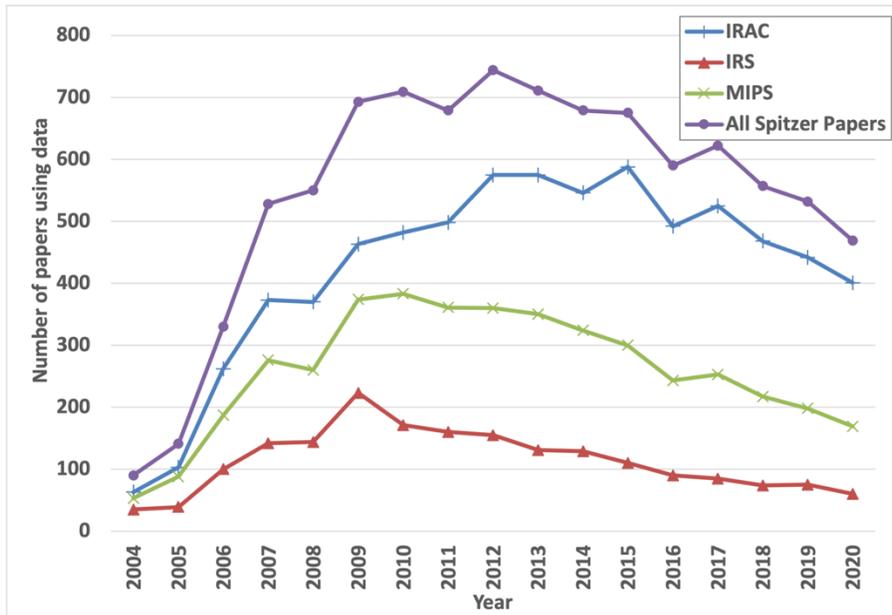

Figure 8: Total number of *Spitzer* papers by instrument. MIPS and IRS were decommissioned in May 2009 and IRAC in Jan 2020. A version of this figure was published as Figure 7 in Scire et al. 2020. Data from 2020 have been added to the plot shown here.

Although IRS data constitute 40.8% of the hours of data taken in the cryogenic mission (Cryo IRAC was 27.1% and MIPS was 32.1%; Table 1) IRS has by far the fewest number of papers of the three instruments (Figure 8). For the cryogenic mission *Spitzer's* imaging instruments MIPS and cryogenic IRAC have much higher papers per hour of observation rates (0.38 and 0.46 papers per hour of data, respectively) than the spectrograph, IRS (0.13 papers per hour of data; Table 1). Imaging data can contain multiple objects in the same frame, and each object can end up being its own paper. In addition, the imaging data are sometimes used more casually as background images for X-ray or radio contour lines, or as finding charts of an area. Authors also are more likely to grab a photometric point from a science-ready image or catalog downloaded from the archive and add it to the data from another source that is driving the main science point of the paper (e.g., adding IRAC flux densities to a spectral energy distribution). The spectra, on the other hand, are mostly of one object each, and papers that use IRS data tend to be analyzing the spectra and discussing spectral features exclusively. Since the spectra are used in a more focused way than imaging data, it is possible that this is why citations to the IRS fundamental paper are more often included in papers that used IRS data than in papers that used other *Spitzer* instruments (Section 3.1.1). Spectra do tend to provide detailed information about the physics of an object being studied and thus are scientifically valuable, but for *Spitzer* the spectral observations also tend to have lower data reuse rates.

The average observation length for IRS observations was 46.2 minutes, for MIPS it was 41.1 minutes and for cryogenic IRAC it was 34.0 minutes (Table 1). While the IRS observations are slightly longer on average, the difference is not enough to explain the difference in the papers per hour of data rate between IRS and the other cryogenic instruments (especially MIPS), or the difference between cryogenic and Warm warm IRAC.

Warm IRAC has a much lower papers per hours of data rate than any of the cryogenic instruments (0.07 papers per hour of data for data taken before cycle 12 (data taken after Cycle 12 are omitted from this number because observations from the last observing cycles are, at the time of the writing of the current paper, still producing their initial papers: Figure 4). It may be in part because the average IRAC observation length increased from 34.0 minutes in the cryogenic mission to 44.7 minutes in the warm mission (Table 1). But it is also due to the nature of science that was selected for the warm mission. See Scire et al. 2020, and Sections 4.4.2, 4.4.3 and 4.4.4 which discuss this in more detail.

### 4.4.2 Program Types

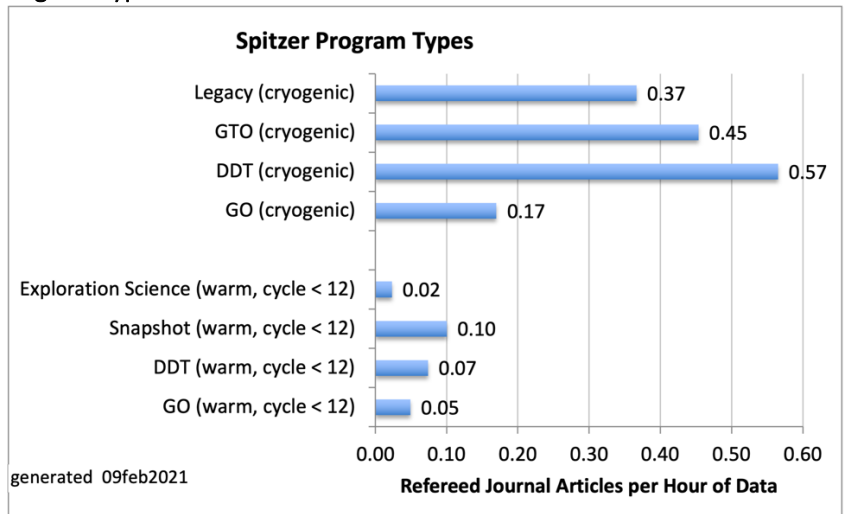

Figure 9: Papers per hour of observation rates sorted by program types for the cryogenic and warm missions. GTO and Legacy programs were cryogenic mission only, and Exploration Science and Snapshot programs were warm mission only. The warm mission only includes data for observing cycles < 12 because later cycles are still in the process of producing their initial papers (Figure 4). See Figure 1 for a breakdown of program types and number of hours of observations in each category. See text for discussion.

Blecksmith et al. 2005 found that for *Chandra*, the GTO programs regularly produced more papers per kilosecond of observations than the GO programs and attributed it to the fact that GTO observers work in large teams, while GO observers tend be one or two people. This is true for *Spitzer* as well, but we believe there are additional contributing factors to the differing rates for *Spitzer* (Figure 9). Before the launch, the Legacy and GTO programs were selected before the GO programs. Because they were selecting targets first, the high-value, easy to observe, very popular fields ended up clustered in the Legacy and GTO programs (with the Legacy programs covering the large fields and the GTO programs covering the small ones). In addition, the Legacy and GTO programs were funded over a span of years, allowing the teams to have a dedicated stable workforce. They also received data analysis funding that was at a higher level than most of the GO programs[12]. All of these factors can help to explain why the papers per hour of observation rate is smaller for the GO programs in the cryogenic mission.

The highest rate of papers produced per hour of data obtained is found for the cryogenic DDT programs. The DDT programs were submitted on a rolling basis throughout the cryogenic and warm missions for emerging science (high risk/high reward) or unanticipated time critical observations. These observations tended to be of exciting events, and were given a proprietary period of at most 90 days. Both of these factors may help explain why they have a higher papers per hour of data rate, but it may also be a function of the small sample size. In total, only 3.3% (1261 hours) of the observed science time in the cryogenic mission was in DDT observations. The rest of the time was in Legacy programs (24.0%), GTOs (17.8%) and the GO programs (54.7%).

---

[12] Cryogenic GO programs were funded according to a formula that was instrument dependent. IRAC observations were funded at a lower level than IRS and MIPS observations, and the mix of instruments used could result in GO funding levels that were higher on a dollars/hour of observation rate than the GTO/Legacy programs, but frequently were not.

For the warm mission, the number of papers per hour of observation rate is significantly lower than the cryogenic mission rates. In particular the rate for the Exploration Science[6] programs is lower, due largely to the nature of the science that warm IRAC was good at doing. With only one instrument operating with two channels, 3.6 and 4.5 microns, it turned out that most of the science selected was extragalactic surveys, extrasolar planet transit/eclipse light curves, extrasolar planet and brown dwarf phase curves and other time series observations (Scire et al. 2020 Figure 8). These kinds of science require a greater time investment for science returns (for example, hundreds of hours invested in a time series may only produce one paper.) Also, in the warm mission, more time was devoted to large Exploration Science surveys, which reduced the number of GO programs (Figure 1). See below (Figure 11 and Figure 12) for a breakdown of how the average observation length and the total time observed impact the papers per hour of observation rate.

The DDT observations in the warm mission did not produce a higher papers per hour of data rate, unlike the cryogenic DDT observations. The warm mission DDT observations do not suffer from as much of a small sample size problem: 9.2% (7791.3 hours) of the observed time in the warm mission was in DDT observations. The average amount of observing time in a cryogenic DDT program was 8.5 hours, while for the warm DDT programs it was 39.6 hours. 62% of the warm DDT time went to programs with more than 100 hours of observations, whereas the largest cryogenic DDT program was allocated 59.3 hours.

### 4.4.3 Legacy/Exploration Science Data Usage

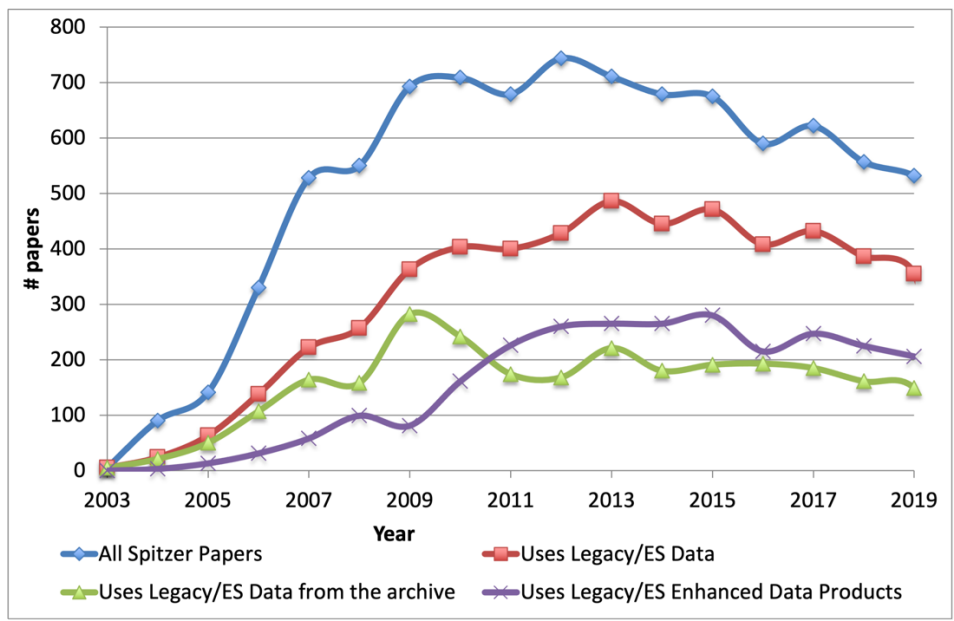

Figure 10. Number of *Spitzer* papers that use data from an Exploration Science (ES) or Legacy program plotted compared to the number of all the *Spitzer* papers. The 'Uses Legacy/ES Data from the archive' line consists of papers that used data downloaded from the *Spitzer* data archive. 'Uses Legacy/ES Enhanced Data Products' papers explicitly state that the enhanced products that were produced by the Legacy/ES teams were used. 'Uses Legacy/ES Data' is the sum of the two previous categories. Papers that used Legacy/ES data could have also used other *Spitzer* data.

The first Legacy program enhanced data products were released in October of 2004, yet the number of papers using these data only grew slowly until 2010, when their usage starts to accelerate (Figure 10). Since it takes $1-3$ years for data to be published after the authors have access to them (Figure 5) that means there was an additional, significant lag in usage of the enhanced data products. It may be that it took time for astronomers to learn of these products and for their confidence in the products to grow enough for them to be preferred over the pipeline products from the Spitzer Science Center pipelines or their own reductions. They also could have been better advertised. In email blasters sent to the astronomical community, the Legacy enhanced products were mentioned two times in total before 2008. In 2008 they were mentioned three times, in 2009 nine times and after 2009 there were regular mentions 4+ times a year. They were also initially only available from a special website that was separate from the rest of the data archive, so the data users had to know to go look for them. As time progressed, they were better integrated into the rest of the archival interfaces at the Spitzer Science Center and IRSA, which has helped their popularity to grow.

To encourage the use of enhanced data products, we suggest that they be placed in a well-advertised and maintained archive, and that the astronomical community be reminded of their existence on a regular basis. We also recommend they be integrated into object search engines (e.g. NASA/IPAC Extragalactic Database (https://ned.ipac.caltech.edu/, etc.) where possible.

### 4.4.4 Program and Observation Length

In Figure 11 and Figure 12, each of the points on a graph are for one Program ID (PID). A few observing programs (Original Legacy Science, and Exploration Science programs, and microlensing programs) were split across multiple PIDs to make bookkeeping for the program easier (e.g., SWIRE (Lonsdale et al. 2004) was split by the field on the sky, the Carnegie Hubble Program (Freedman et al. 2008) was split by the target type, and microlensing (Gould et al. 2014) was split by the week observed). Only PIDs where >75% of the hours of observations were completed are included (1573 cryogenic PIDs and 507 warm PIDs). A total of 91 PIDs were excluded; 45 of the excluded PIDs were Cycle 5 priority 3 programs that were not completed before the cryogen was exhausted. The cryogenic depletion date was somewhat unknown, so enough science was selected to fill the time to the latest possible date, with the priority 1 and 2 science scheduled before the priority 3 science. The warm mission Cycles 12-14 have been omitted from these graphs because the data are still fairly new, and observers are still working on the first publications (see Figure 4).

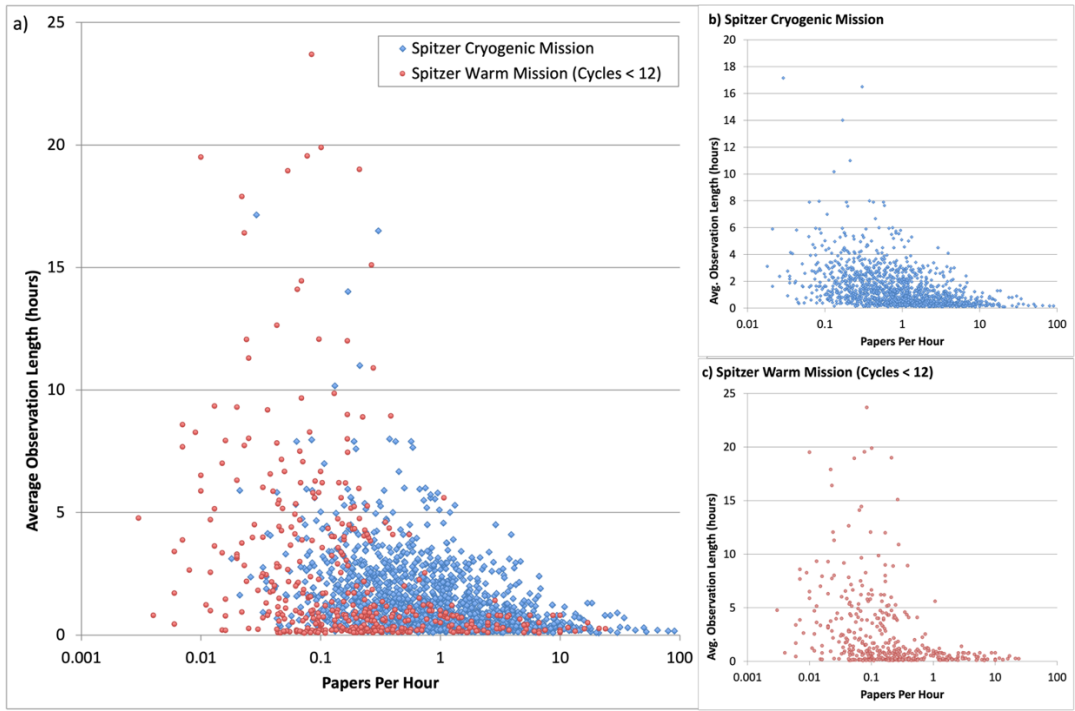

Figure 11: The average observation length versus the number of refereed journal articles per hour of data observed for that PID. The horizontal axis is logarithmic. a) all *Spitzer* PIDs for all the observing cycles before Cycle 12, b) just the cryogenic mission c) just the warm mission for programs before Cycle 12. The scale of the axes on all the three graphs is different. PIDs with a very large papers per hour of observation rates tend to have modest allocations (< 1 hour) with short individual observations.

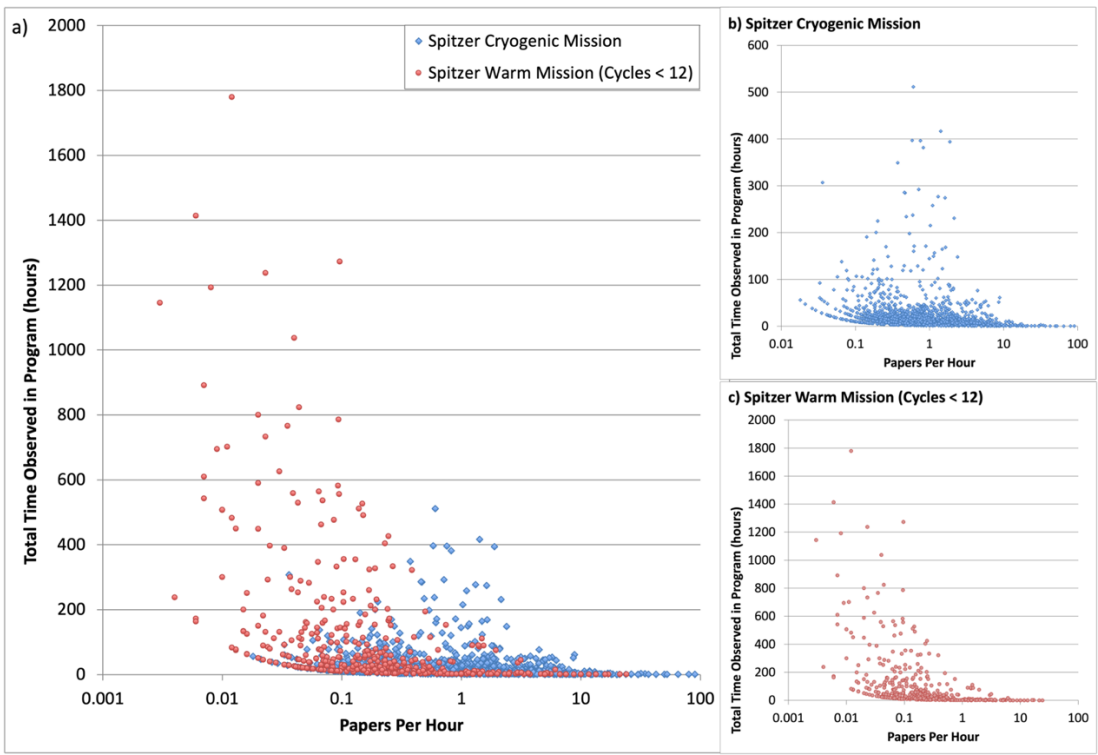

Figure 12: The total time observed versus the number of refereed journal articles per hour of data observed taken for that PID. The horizontal axis is logarithmic. a) all the *Spitzer* PIDs for all the observing cycles before Cycle 12, b) just the cryogenic mission c) just the warm mission for programs before Cycle 12. The scale of the axes on all the three graphs is different. PIDs with very large numbers of papers per hour of observation observed rates tend to have modest allocations (< 1 hour) with short individual observations

By looking at Figure 11 and Figure 12, it is clear that PIDs with large numbers of papers per hour of observation rates tend to have modest allocations (< 1 hour) with short individual observations. Longer observations tend to produce fewer papers per hour of observation. For *Spitzer*, most of the long (> 5 hour) observations were exoplanet transit or eclipse and exoplanet and brown dwarf phase curve measurements. There was a 24-hour limit on an observation's duration, and so frequently for these measurements several long observations were strung together in a row to produce longer data sets. (The longest continuous phase curve observed was 87 hours broken into 4 observations chained together, and the longest observations were 21 and 22 days with breaks only for downlinks and spacecraft calibrations.)

Surprisingly, the average observation length for the entire cryogenic mission is almost identical to that of the entire warm mission: 48.2 minutes and 48.1 minutes, respectively. However, the standard deviation of the observation length is very different for the two mission phases: the cryogenic mission had a standard deviation of 64.6 minutes, while for the warm mission it is 113.9 minutes.

We would like to explicitly point out that the observations that produced the highest papers per hour of data rate were of specific, very popular areas of the sky (for more information on which areas of the sky see Scire et al. 2020). We believe that it is beneficial to the community to make sure that those areas of the sky are observed by most observatories and the data be put into a well-advertised and maintained archive. It is also apparent in Figure 11 and Figure 12 that just because a program has a modest allocation or short observations, it does not necessarily produce a higher papers per hour of data rate. In fact, for the cryogenic mission the programs with the highest allocation (the Legacy programs) have middling rates of papers per hour of data (Figure 12b) that are continuing to increase at rates exceeding that of the smaller programs. Comparing the large cryogenic program to the warm programs of similar allocation (Figure 12a), the cryogenic programs have higher papers per hour of data rates. This may be because the cryogenic data are older and have had more time to be republished, or it may be because the most useful fields on the sky were observed earlier in the mission. Counting papers with this level of detail for another ~ 10 years would answer this question, however there is no funding to do so.

## 5  Summary

The cryogenic and warm *Spitzer* missions had very different rates of publication in refereed journals. After the cryogen was exhausted in May 2009, only half of the IRAC instrument remained operational, and the number of programs approved in each roughly annual observing cycle decreased from ~ 300 to ~ 60. At the same time, the overall budget for the mission (including the data analysis funding distributed to the observers) was reduced. The types of science that best suited the warm IRAC instrument tended to have substantially longer individual observations than those made in the cryogenic mission. All these factors combined led to a reduction in the number of papers that were published annually in the warm mission. However, the fraction of *Spitzer* observing time that was published and re-published did not substantially change between the cryogenic and warm missions. The total numbers of papers from each of the cryogenic and the warm mission do not capture the complexity of what factors affected the publication rate in each phase of the *Spitzer* Mission, so raw paper counts are not good metrics for the "success" of an observatory.

Because the publication rates between the warm and cryogenic missions differ so greatly, we find that for *Spitzer* we have to be careful when comparing the publications from *the same observatory* since the mission profile changed significantly between the warm and cryogenic missions. This also suggests that more care needs to be taken when doing cross-observatory comparisons based on raw counts of paper numbers. We encourage more observatories to consider the fraction of observing time published as a most robust metric, even though it does require more resources to calculate.

Most of the time there is a delay of one to four years between the time when the data are taken, and the time when results from these data are published for the first time. This delay depends on the instrument the data were taken with. The simpler IRAC data were on average published six months faster than the data from the instruments that required more processing (MIPS, IRS). Also, the data from *Spitzer*'s two imaging instruments (IRAC and MIPS) can often be used in multiple, often independent research projects, and therefore result in multiple papers. Data from the spectrograph (IRS) often are taken for a single, targeted object, and therefore are less amenable to be used in large surveys that cover huge swaths of the sky (and that often produce many papers).

The absence of proprietary periods for data contained in several large programs led to a quick upturn in the number of papers that used *Spitzer* data at the start of the mission. Ready-to-use (reduced) data that are stored in a well-advertised, well-organized, well-documented, and freely accessible data archive have contributed to the publication of a large number of archival papers that often use the same data sets multiple times. The enhanced data products from the Legacy and Exploration Science Teams have been very heavily used, and the number of hours of data published by papers that are using the entire catalogue/mosaics from the Legacy/Exploration Science programs far exceeds any other usage of the data. However, it took some time for these teams to make their products available and make other researchers aware of the value of these data. A large fraction of *Spitzer* papers use data from a small fraction of all the executed observing programs, specifically utilizing the Legacy, GTO, or Exploration Science programs. The original Legacy and GTO programs were given target selection priority before the launch. For this reason, many of the interesting and most obvious targets were included in those programs, and those data are frequently re-used.

A long operational lifetime of an observatory and/or an instrument allows observers to expand their sample sizes and to acquire new epochs of observations, producing more publications and increasing the reuse of the archival data.

Smaller program allocations produce many papers very soon after observation, whereas the larger programs result in a longer ramp up time but have a longer, more robust publications tail. For Spitzer, the mix of small and large programs in the cryogenic mission produced swift upslopes in the number of papers produced and a robust long term publication rate and data sets that are still in heavy use over a decade after the cryogenic mission completed.

The changes made for the warm mission allowed for a new investigative parameter space and allowed observers to answer questions that require significantly more investment of observing time. While the paper counts may be lower, the warm mission allowed significantly more exoplanet light curves to be taken with much longer durations, provided confirmation observations for *Kepler* (Koch et al. 2010) and *TESS* (Ricker et al. 2015), helped refine the Hubble Constant, assisted in the discovery of the seven Earth sized planets orbiting TRAPPIST-1 (Gillon et al. 2016), inspired a Google Doodle and a front page of the New York Time (above the fold) (Chang 2017), and then spent hundreds of hours studying the TRAPPIST-1 system to make it the best categorized solar system outside our own.

Finally, we urge authors of journal papers to cite the observatory fundamental papers, and to provide enough information in their publications to allow a unique identification of the data that were used. We also urge observatories and observers to pick names for their observing programs that are unique enough to allow a full-text search and to identify the relevant results easily. Since 2012, only 62% of papers that used data from Spitzer cited either the fundamental papers or the observatory acknowledgement statement. About 15% of the papers did not supply enough information to uniquely identify the Spitzer data used in the publication. Given the time delay from data taking to the publication of the results, and the huge program sizes and resulting data sets that were taken towards the end of the *Spitzer* mission, and if funding existed, it would be interesting to count *Spitzer* papers with the same careful analysis five to seven years after the end of the mission to determine what has happened to the *Spitzer* warm mission data publication rates.

# 6 Acknowledgements

Special thanks to people who have helped with this project throughout the mission: Ben Chan, Jessica Krick, Aomawa Shields, Nancy Silbermann, and to the entire *Spitzer Space Telescope* Project. Working on *Spitzer* has been one of the great joys of our lives. We would also like to thank the anonymous referee whose comments improved the presentation in this paper. This work is based on observations made with the *Spitzer Space Telescope*, which was operated by the Jet Propulsion Laboratory, California Institute of Technology under a contract with NASA. This research has made use of NASA's Astrophysics Data System. This research has made use of the NASA/IPAC Infrared Science Archive, which is funded by the National Aeronautics and Space Administration and operated by the California Institute of Technology. Ironically, since this paper does not use data taken during science observations as a primary source, it would not be included in the paper counts referenced above.